\documentclass[english]{llncs}
\usepackage[T1]{fontenc}
\usepackage[latin9]{inputenc}
\usepackage{rotfloat}
\usepackage{amsmath}
\usepackage{graphicx}

\makeatletter

\providecommand{\tabularnewline}{\\}

\makeatother

\usepackage{babel}
\begin{document}

\title{A reliability-based approach for influence maximization using the
evidence theory}

\author{Siwar Jendoubi\inst{1} and Arnaud Martin\inst{2}}

\institute{LARODEC, University of Tunis, ISG Tunis, Avenue de la Liberté, Cité
Bouchoucha, Le Bardo 2000, Tunisia\\
jendoubi.siwar@yahoo.fr \and  DRUID, IRISA, University of Rennes
1, Rue E. Branly, 22300 Lannion, France \\
Arnaud.Martin@univ-rennes1.fr}
\maketitle
\begin{abstract}
The influence maximization is the problem of finding a set of social
network users, called influencers, that can trigger a large cascade
of propagation. Influencers are very beneficial to make a marketing
campaign goes viral through social networks for example. In this paper,
we propose an influence measure that combines many influence indicators.
Besides, we consider the reliability of each influence indicator and
we present a distance-based process that allows to estimate the reliability
of each indicator. The proposed measure is defined under the framework
of the theory of belief functions. Furthermore, the reliability-based
influence measure is used with an influence maximization model to
select a set of users that are able to maximize the influence in the
network. Finally, we present a set of experiments on a dataset collected
from Twitter. These experiments show the performance of the proposed
solution in detecting social influencers with good quality.

\keywords{Influence measure, influence maximization, theory of belief functions, reliability, social network.}  
\end{abstract}

\section{Introduction}

The influence maximization problem has attracted a great attention
in these last years. The main purpose of this problem is to find a
set of influence users, $S$, that can trigger a large cascade of
propagation. These users are beneficial in many application domains.
A well-known application is the viral marketing. Its purpose is to
promote a product or a brand through viral propagation through social
networks. Then, several research works were introduced in the literature
\cite{Kempe03,Goyal12,Aslay14,Baghmolaei} trying to find an optimal
set of influence users in a given social network. However, the quality
of the detected influence users stills always an issue that must be
resolved. 

The problem of identifying influencers was first modeled as a learning
problem by Domingos and Richardson \cite{Domingos01} in 2001. Furthermore,
they defined the customer's network value,\textit{\scriptsize{} }\textit{i.e.}
``the expected profit from sales to other customers he may influence
to buy, the customers those may influence, and so on recursively''
\cite{Domingos01}. Moreover, they considered the market to be a social
network of customers. Later in 2003, Kempe \textit{et al.} \cite{Kempe03}
formulated the influence problem as an optimization problem. Indeed,
they introduced two influence maximization models: the \textit{Independent
Cascade Model} (ICM) and the \textit{Linear Threshold Model} (LTM).
These models estimate the expected propagation size, $\sigma_{M}$,
of a given node or set of nodes through propagation simulation models.
Besides, \cite{Kempe03} proved the NP-Hardness of the maximization
of $\sigma_{M}$. Then, they proposed the greedy algorithm to approximate
the set of nodes that maximizes $\sigma_{M}$. ICM and LTM just need
the network structure to select influencers. However, these solutions
are shown in \cite{Goyal12} to be inefficient to detect good influencers.

When studying the state of the art of the influence maximization problem,
we found that most of existing works use only the structure of the
network to select seeds. However, the position of the user in the
network is not sufficient to confirm his influence. For example, he
may be a user that was active in a period of time, then, he collected
many connections, and now he is no longer active. Hence, the user's
activity is an interesting parameter that must be considered while
looking for influencers. Besides to the user's activity in the network,
many other important influence indicators are not considered. Among
these indicators, we found the sharing and tagging activities of network
users. These activities allow the propagation of social messages from
one user to another. Also, the tagging activity is a good indicator
of the user's importance in the network. In fact, more he is tagged
in others' posts, more he is important for them. Therefore, considering
such influence behaviors will be very beneficial to improve the quality
of selected seeds.

To resolve the influencers quality issue, many influence indicators
must be used together to characterize the influence that exerts one
user on another \cite{Jendoubi2017}. An influence indicator may be
the number of neighbors, the frequency of posting in the wall, the
frequency of neighbor\textquoteright{}s likes or shares, \textit{etc}.
Furthermore, a refined influence measure can be obtained through the
fusion of two or more indicators. A robust framework of information
fusion and conflict management that may be used in such a case is
the framework of the theory of belief functions \cite{Shafer76}.
Indeed, this theory provides many information combination tools that
are shown to be efficient \cite{Dempster67a,Smets94a} to combine
several pieces of information having different and distinct sources.
Other advantages of this theory are about uncertainty, imprecision
and conflict management. 

In this paper, we tackle the problem of influence maximization in
a social network. More specifically, our main purpose is to detect
social influencers with a good quality. For this goal, we introduce
a new influence measure that combines many influence indicators and
considers the reliability of each indicator to characterize the user's
influence. The proposed measure is defined through the theory of belief
functions. Another important contribution in the paper is that we
use the proposed influence measure for influence maximization purposes.
This solution allows to detect a set of influencers having a good
quality and that can maximize the influence in the social network.
Finally, a set of experiments is made on real data collected from
Twitter to show the performance of the proposed solution against existing
ones and to study the properties of the proposed influence measure. 

This paper is organized as follows: related works are reviewed in
section 2. Indeed, we present some data-based works and existing evidential
influence measures. Section 3 presents some basic
concepts about the theory of belief functions. Section 4 is dedicated
to explaining the proposed reliability-based influence measure. Section
5 presents a set of experiments showing the efficiency of the proposed
influence measure. Finally, the paper is concluded in Section 6.

\section{Related works}

The influence maximization is a relatively new research problem. Its
main purpose is to find a set of $k$ social users that are able to
trigger a large cascade of propagation through the word of mouth effect.
Since its introduction, many researchers have turned to this problem
and several solutions are introduced in the literature \cite{Kempe03,Kempe2005,Goyal12,Chen12,Jendoubi2016a}.
In this section, we present some of these works.

\subsection{Influence maximization models}

The work of Kempe \textit{et al.} \cite{Kempe03} is the first to
define the problem of finding influencers in a social network as a
maximization problem. In fact, they defined the influence of a given
user or set of users, $S$, as the expected number of affected nodes,
$\sigma_{M}\left(S\right)$, \textit{i.e.} nodes that received the
message. Furthermore, they estimated this influence through propagation
simulation models which are the \textit{Independent Cascade Model}
(ICM) and the \textit{Linear Threshold Model} (LTM). Next, they used
a greedy-based solution to approximate the optimal solution. Indeed,
they proved the NP-Hardness of the problem.

In the literature, many works were conducted to improve the running
time when considering ICM and LTM. Leskovec\textit{ et al.} \cite{Leskovec07b}
introduced the \textit{Cost Effective Lazy Forward} (CELF) algorithm
that is proved to be 700 times faster than the solution of \cite{Kempe03}.
 Kimura and Saito \cite{Kimura2006} proposed the \textit{Shortest-Path
Model (SPM)} which is a special case of the ICM. Bozorgi\textit{ et
al.} \cite{Bozorgi2016} considered the community structure, \textit{i.e.}
a community is a set of social network users that are connected more
densely to each other than to other users from other communities \cite{Mumu2014,Zhouab2015},
in the influence maximization problem. 

The \textit{Credit Distribution model} (CD) \cite{Goyal12} is an
interesting solution that investigates past propagation to select
influence users. Indeed, it uses past propagation to associate to
each user in the network an influence credit value. The influence
spread function is defined as the total influence credit given to
a set of users $S$ from the whole network. The algorithm scans the
data (past propagation) to compute the total influence credit of a
user $v$ for influencing its neighbor $u$. In the next step, the
CELF algorithm \cite{Leskovec07b} is run to approximate the set of
nodes that maximizes the influence spread in the network.

\subsection{Influence and theory of belief functions}

The theory of belief functions was used to measure the user's influence
in social networks. In fact, this theory allows the combination of
many influence indicators together. Besides, it is useful to manage
uncertainty and imprecision. This section is dedicated to present
a brief description of existing works that use the theory of belief
functions for measuring or maximizing the influence.

An evidential centrality (EVC) measure was proposed by Wei \textit{et
al.} \cite{Wei13} and it was used to estimate the influence in the
network. EVC is obtained through the combination of two BBAs defined
on the frame $\left\{ high,\, low\right\} $. The first BBA defines
the evidential degree centrality and the second one defines the evidential
strength centrality of a given node. A second interesting, work was
also introduced to measure the evidential influence, it is the work
of \cite{Gao13}. They proposed a modified EVC measure. It considers
the actual node degree instead of following the uniform distribution.
Furthermore, they proposed an extended version of the semi-local centrality
measure \cite{Chen12} for weighted networks. Their evidential centrality
measure is the combined BBA distribution of the modified semi-local
centrality measure and the modified EVC. The works of \cite{Gao13}
and \cite{Wei13} are similar in that, they defined their measures
on the same frame of discernment, they used the network structure
to define the influence.

Two evidential influence maximization models are recently introduced
by Jendoubi \textit{et al.} \cite{Jendoubi2017}. They used the theory
of belief functions to estimate the influence that exerts one user
on his neighbor. Indeed, their measure fuses several influence indicators
in Twitter like the user's position in the network, the user's activity,
\textit{etc}. This paper is based on our previous work \cite{Jendoubi2017}.
However, the novelty of this paper is that we not only combine many
influence indicators to estimate the user's influence, but also we
consider the reliability of each influence indicator in characterizing
the influence.

\section{Theory of belief functions}

In this section, we present the \textit{theory of belief functions},
also called \textit{evidence theory} or \textit{Dempster-Shafer theory}.
It was first introduced by Dempster \cite{Dempster67a}. Next, the
mathematical framework of this theory was detailed by Shafer in his
book \textit{``A mathematical theory of evidence}'' \cite{Shafer76}.
This theory is used in many application domains like pattern clustering
\cite{Denoeux2016,Liu2015} and classification \cite{Liu2016,Jendoubi14a,Jendoubi2015}.
Furthermore, this theory is used for analyzing social networks and
measuring the user's influence \cite{Gao13,Wei13,Jendoubi2017}.

Let us, first, define the \textit{frame of discernment} which is the
set of all possible decisions: 
\begin{equation}
\Omega=\left\{ d_{1},d_{2},...,d_{n}\right\} 
\end{equation}
The \textit{mass} function, also called \textit{basic belief assignment}
(BBA), $m^{\Omega}$, defines the source's belief on $\Omega$ as
follows: 
\begin{eqnarray}
2^{\Omega} & \rightarrow & \left[0,1\right]\nonumber \\
A & \mapsto & m\left(A\right)
\end{eqnarray}
such that $2^{\Omega}=\left\{ \emptyset,\left\{ d_{1}\right\} ,\left\{ d_{2}\right\} ,\left\{ d_{1},d_{2}\right\} ,...,\left\{ d_{1},d_{2},...,d_{n}\right\} \right\} $.
The set $2^{\Omega}$ is called \textit{power set}, \textit{i.e.}
the set of all subsets of $\Omega$. The value assigned to the subset
$A\subseteq\Omega$, $m\left(A\right)$, is interpreted as the source's
support or belief on $A$. The BBA distribution, $m$, must respect
the following condition: 
\begin{equation}
\sum_{A\subseteq\Omega}m\left(A\right)=1\label{eq:mass}
\end{equation}
We call $A$ \textit{focal element} of $m$ if we have $m(A)>0$.
The \textit{discounting} procedure allows to consider the \textit{reliability}
of the information source. Let $\alpha\in\left[0,1\right]$ be our
reliability on the source of the BBA $m$, then the discounted BBA
$m^{\alpha}$ is obtained as follows:

\begin{equation}
\begin{cases}
m^{\alpha}\left(A\right) & =\alpha.m\left(A\right),\,\forall A\in2^{\Omega}\setminus\left\{ \Omega\right\} \\
m^{\alpha}\left(\Omega\right) & =1-\alpha.\left(1-m\left(\Omega\right)\right)
\end{cases}\label{eq:discounting}
\end{equation}

The information fusion is important when we want to fuse many influence
indicators together in order to obtain a refined influence measure.
Then, the theory of belief functions presents several combination
rules. The\textbf{ }\textit{Dempster's rule} of combination \cite{Dempster67a}
is one of these rules. It allows to combine two distinct BBA distributions.
Let $m_{1}$ and $m_{2}$ be two BBAs defined on $\Omega$, Dempster's
rule is defined as follows: 
\begin{equation}
m_{1\oplus2}\left(A\right)=\begin{cases}
\frac{{\displaystyle \sum_{B\cap C=A}}m_{1}\left(B\right)m_{2}\left(C\right)}{1-{\displaystyle \sum_{B\cap C=\emptyset}}m_{1}\left(B\right)m_{2}\left(C\right)}, & A\subseteq\Omega\setminus\left\{ \emptyset\right\} \\
0 & if\, A=\emptyset
\end{cases}\label{eq:dempster rule}
\end{equation}

In the next section, we present some relevant existing influence measures
and influence maximization models.

\section{Reliability-based influence maximization}

In this paper, we propose an influence measure that fuses many influence
indicators. Furthermore, we assume that these indicators may do not
have the same reliability in characterizing the influence. Then, some
indicators may be more reliable than the others. In this section,
we present the proposed reliability-based influence measure, the method
we use to estimate the reliability of each indicator and the influence
maximization model we use to maximize the influence in the network.

\subsection{Influence characterization}

Let $G=\left(V,E\right)$ be a social network, where $V$ is the set
of nodes such that $u,v\in V$ and $E$ is the set of links such that
$\left(u,v\right)\in E$. To estimate the amount of influence that
exerts one user, $u$, on his neighbor, $v$, we start first by defining
a set of influence indicators, $I=\left\{ i_{1},i_{2},\ldots,i_{n}\right\} $
characterizing the influence. These indicators may differ from a social
network to another. We note that we are considering quantitative indicators.
Let us take Twitter as example, we can define the following three
indicators: 1) the number of common neighbors between $u$ and $v$,
2) the number of times $v$ mentions $u$ in a tweet, 3) the number
of times $v$ retweets from $u$.

In the next step, we compute the value of each defined indicator for
each link $\left(u,v\right)$ in the network. Then, $\left(u,v\right)$
will be associated with a vector of values. In a third step, we need
to normalize each computed value to the range $\left[0,1\right]$.
This step is important as it puts all influence indicators in the
same range.

In this stage, we have a vector of values of the selected influence
indicators: 
\begin{equation}
W_{\left(u,v\right)}=\left(i_{\left(u,v\right)_{1}}=w_{1},i_{\left(u,v\right)_{2}}=w_{2},\ldots,i_{\left(u,v\right)_{n}}=w_{n}\right)
\end{equation}
The elements of $W_{\left(u,v\right)}$ are in the range $\left[0,1\right]$,
\textit{i.e.} $w_{1},w_{2},\ldots w_{n}\in\left[0,1\right]$, and
we define a vector $W_{\left(u,v\right)}$ for each link $\left(u,v\right)$
in the network. Next, we estimate a BBA for each indicator value and
for each link. Then, if we have $n$ influence indicators, we will
obtain $n$ BBA to model each of these indicators for a given link.
Let us first, define $\Omega=\left\{ I,P\right\} $ to be the frame
of discernment, where $I$ models the influence and $P$ models the
passivity of a given user. For a given link $\left(u,v\right)$ and
a given influence indicator $i_{\left(u,v\right)_{j}}=w_{j}$, we
estimate its BBA  on the fame $\Omega$ as follows:

\begin{eqnarray}
m_{\left(u,v\right)_{j}}\left(I\right) & = & \frac{w_{j}-\min_{\left(u,v\right)\in E}\left(i_{\left(u,v\right)_{j}}\right)}{\max_{\left(u,v\right)\in E}\left(i_{\left(u,v\right)_{j}}\right)-\min_{\left(u,v\right)\in E}\left(i_{\left(u,v\right)_{j}}\right)}\\
m_{\left(u,v\right)_{j}}\left(P\right) & = & \frac{\max_{\left(u,v\right)\in E}\left(i_{\left(u,v\right)_{j}}\right)-w_{j}}{\max_{\left(u,v\right)\in E}\left(i_{\left(u,v\right)_{j}}\right)-\min_{\left(u,v\right)\in E}\left(i_{\left(u,v\right)_{j}}\right)}
\end{eqnarray}
After this step, the influence that exerts a user $u$ on his neighbor
$v$ is characterized by a set of $n$ influence BBAs. In the next
section, we present the method we use to estimate the reliability
of each defined BBA.

\subsection{Estimating reliability\label{sub:Estimating-reliability}}

The selected influence indicators may do not have the same reliability
in characterizing the user's influence. Then, we estimate the reliability,
$\alpha_{j}$, of each influence indicator. We assume that ``\textit{the
farthest from the others the indicator is, the less reliable it is}''.
For that purpose, we follow the approach introduced by Martin \textit{et
al.} \cite{Martin08a} to estimate reliability. Besides, we note that
this operator considers our assumption. In this section we detail
the steps of \cite{Martin08a} operator we used to estimate the reliability
of each influence indicator in this paper. 

Let us consider the link $\left(u,v\right)$, we have a set of $n$
BBAs to characterize the chosen influence indicators, $\left(m_{\left(u,v\right)_{1}},m_{\left(u,v\right)_{2}},\ldots,m_{\left(u,v\right)_{n}}\right)$.
Our purpose is to estimate the reliability of each indicator against
the others. To estimate the reliability, $\alpha_{j}$, of the BBA
$m_{\left(u,v\right)_{j}}$, we start by computing the distance between
$m_{\left(u,v\right)_{j}}$ and each BBA from the rest of $n-1$ BBAs
that characterizes the influence of $u$ on $v$, \textit{i.e.} $\left(m_{\left(u,v\right)_{1}},m_{\left(u,v\right)_{2}},\ldots,m_{\left(u,v\right)_{j-1}},m_{\left(u,v\right)_{j+1}},\ldots,m_{\left(u,v\right)_{n}}\right)$:

\begin{equation}
\delta_{i}^{j}=\delta(m_{j},m_{i})
\end{equation}
To estimate these distances, we can use the \textit{Jousselme} distance
\cite{Jousselme01a} as follows:
\begin{equation}
\delta\left(m_{j},m_{i}\right)=\sqrt{\frac{1}{2}\left(m_{j}-m_{i}\right)^{T}\underset{=}{D}\left(m_{j}-m_{i}\right)}
\end{equation}
such that $\underset{=}{D}$ is an $2^{N}\times2^{N}$ matrix, $N=|\Omega|$
and $D\left(A,B\right)=\frac{|A\cap B|}{|A\cup B|}$.

Next, we compute the average of all obtained distance values as follows:

\begin{equation}
C_{j}=\frac{\delta_{j}^{1}+\delta_{j}^{2}+\ldots+\delta_{j}^{n}}{n-1}
\end{equation}
such that $\left(\delta_{1},\delta_{2},\ldots,\delta_{n}\right)$
are the distance values between $m_{\left(u,v\right)_{j}}$ \\
and $\left(m_{\left(u,v\right)_{1}},m_{\left(u,v\right)_{2}},\ldots,m_{\left(u,v\right)_{n}}\right)$,
$\left(\delta\left(m_{j},m_{j}\right)=0\right)$. We use the average
distance $C_{j}$ to estimate the reliability, $\alpha_{j}$, of the
$j^{\textrm{th}}$ influence indicator in characterizing the influence
of $u$ on $v$ as follows:

\begin{equation}
\alpha_{j}=f\left(C_{j}\right)
\end{equation}
where $f$ is a decreasing function. The function $f$ can be defined
as \cite{Martin08a}:

\begin{equation}
\alpha_{j}=\left(1-\left(C_{j}\right)^{\lambda}\right)^{1/\lambda}\label{eq:discount}
\end{equation}
where $\lambda>0$. 

After applying all these steps, we obtain the estimated value of the
BBA reliability, $\alpha_{j}$. To consider this reliability, we apply
the discounting procedure described in equation (\ref{eq:discounting}).
Then, we apply these steps for all defined BBAs on every link in the
network.

\subsection{Influence estimation}

After discounting all BBAs of each link in the network, we use them
to estimate the influence that exerts one user on his neighbor. For
this purpose, let us consider the link $\left(u,v\right)$ and its
discounted set of BBAs $\left(m_{\left(u,v\right)_{1}}^{\alpha_{1}},m_{\left(u,v\right)_{2}}^{\alpha_{2}},\ldots,m_{\left(u,v\right)_{n}}^{\alpha_{n}}\right)$.
We define the global influence BBA that exerts $u$ on $v$ to be
the BBA  that fuses all discounted BBAs defined on $\left(u,v\right)$.
For this aim, we use the Dempster's rule of combination (see equation
(\ref{eq:dempster rule})) to combine all these BBAs as follows:

\begin{equation}
m_{\left(u,v\right)}=m_{\left(u,v\right)_{1}}^{\alpha_{1}}\oplus m_{\left(u,v\right)_{2}}^{\alpha_{2}}\oplus\ldots\oplus m_{\left(u,v\right)_{n}}^{\alpha_{n}}
\end{equation}
The BBA distribution $m_{\left(u,v\right)}$ is the result of this
combination. 

Consequently, we define the influence that exerts $u$ on $v$ to
be the amount of belief given to $\left\{ I\right\} $ as:

\begin{equation}
Inf\left(u,v\right)=m_{\left(u,v\right)}\left(I\right)
\end{equation}
The novelty of this evidential influence measure is that it considers
several influence indicators in a social network and it takes into
account the reliability of each defined indicator against the others.
Our evidential influence measure can be considered as a generalization
of the evidential influence measure introduced in the work of Jendoubi
\textit{et al.} \cite{Jendoubi2017}.

To maximize the influence in the network, we need to define the amount
of influence that exerts a set of nodes, $S$, on the hole network.
It is the total influence given to $S$ for influencing all users
in the network. Then, we estimate the influence of $S$ on a user
$v$ as follows \cite{Jendoubi2017}:

\begin{equation}
Inf\left(S,v\right)=\begin{cases}
1 & if\, v\in S\\
{\displaystyle \sum_{u\in S}}{\displaystyle \sum_{x\in IN\left(v\right)\cup v}}Inf\left(u,x\right).Inf\left(x,v\right) & Otherwise
\end{cases}
\end{equation}
where $Inf\left(v,v\right)=1$ and $IN\left(v\right)$ is the set
of in-neighbors of $v$, \textit{i.e.} if $\left(u,v\right)$ is a
link in the network then $u$ is an in-neighbor of $v$. Next, we
define the influence spread function that computes the amount of influence
of $S$ on the network as follows:

\begin{equation}
\sigma\left(S\right)=\sum_{v\in V}Inf\left(S,v\right)
\end{equation}
To maximize the influence that exerts a set of users $S$ on the network,
we need to maximize $\sigma\left(S\right)$,\textit{ i.e.} $\underset{S}{\textrm{argmax}}\,\sigma\left(S\right)$.
The influence maximization under the evidential model is demonstrated
to be NP-Hard. Furthermore, the function, $\sigma\left(S\right)$,
is monotone and submodular. Proof details can be found in \cite{Jendoubi2017}.
Consequently, a greedy-based solution can perform a good approximation
of the optimal influence users set $S$. In such cases, the cost effective
lazy-forward algorithm (CELF) \cite{Leskovec07b} is an adaptable
maximization algorithm. Besides, it needs only two passes of the network
nodes and it is about 700 times faster than the greedy algorithm.
More details about CELF-based solution used in this paper can be found
in \cite{Jendoubi2017}.

After the definition of the reliability-based evidential influence
measure and the influence spread function, we move to the experiments.
Indeed, we made a set of experiments on real data to show the performance
of our solution.

\section{Results and discussion}

This section is dedicated to the experiments. In fact, we crawled
Twitter data for the period between the 08-09-2014 and 03-11-2014.
Table \ref{tab:Statistics} presents some statistics of the dataset. 

\begin{table}
\caption{Statistics of the data set\label{tab:Statistics} \cite{Jendoubi2017}}

\begin{centering}
\begin{tabular}{|c|c|c|c|c|}
\hline 
Nbr of users  & Nbr of tweets  & Nbr of follows  & Nbr of retweets  & Nbr of mentions\tabularnewline
\hline 
36274  & 251329  & 71027  & 9789  & 20300\tabularnewline
\hline 
\end{tabular}
\par\end{centering}

\end{table}

To characterize the influence users on Twitter, we choose the following
three influence indicators: 1) the number of common neighbors between
$u$ and $v$, 2) the number of times $v$ mentions $u$ in a tweet,
3) the number of times $v$ retweets from $u$. Next, we apply the
process described above in order to estimate the amount of influence
that exerts each user $u$ on his neighbor $v$ in the network.

To evaluate the proposed reliability-based solution, we compare the
proposed solution to the evidential model of Jendoubi \textit{et al.}
\cite{Jendoubi2017}. Furthermore, we choose four comparison criteria
to compare the quality of the detected influence users by each experimented
model. Those criteria are the following: 1) number of accumulated
follow, 2) number of accumulated mention, 3) number of accumulated
retweet, 4) number of accumulated tweet. Indeed, we assume that an
influence user with a good quality is a highly followed user, mentioned
and retweeted several times and active in terms of tweets.

\begin{sidewaysfigure}
\begin{centering}
\includegraphics[scale=0.5]{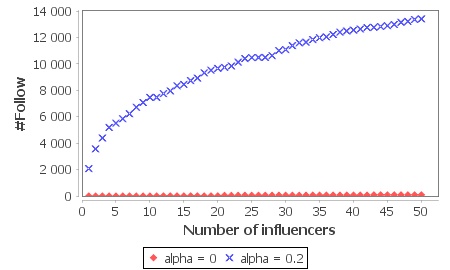}
\includegraphics[scale=0.5]{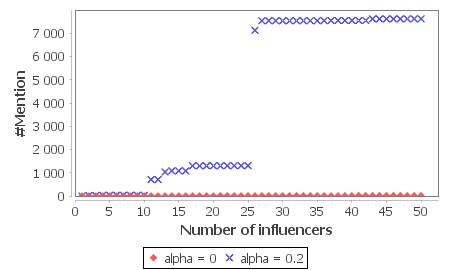}
\par\end{centering}

\begin{centering}
\includegraphics[scale=0.5]{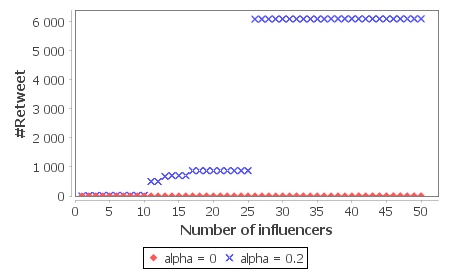}
\includegraphics[scale=0.5]{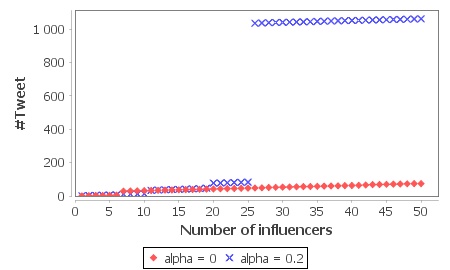}
\par\end{centering}

\caption{Comparison of the reliability-based evidential model with three $\alpha$
fixed value\label{fig:fixedalpha}}

\end{sidewaysfigure}

In a first experiment, we compare the behavior of the proposed measure
with fixed values of indicator reliability. Figure \ref{fig:fixedalpha}
presents the obtained results for two fixed values of $\alpha_{j}=\alpha$,
which are $\alpha_{j}=\alpha=0$ and $\alpha_{j}=\alpha=0.2$. In
Figure \ref{fig:fixedalpha} we have a comparison according to the
four criteria, \textit{i.e.} \#Follow, \#Mention, \#Retweet and \#Tweet
shown in the y-axis of the sub-figures. We comapre the experimented
measure using the set of selected seed for each value of $\alpha_{j}$.
Besides, we fixe the size of the set $S$ of selected influence users
to $50$ influencers, \textit{i.e.} shown in the x-axis of each sub-figure. 

According to Figure \ref{fig:fixedalpha}, we notice that when $\alpha=0$
(red scatter plots), the proposed reliability-based model does not
detect good influencers according to the four comparison criteria.
In fact, the red scatter plot ($\alpha=0$) is very near to the x-axis
in the case of the four criteria, which means that the detected influencer
are neither followed, nor mentioned, nor retweeted. Besides, they
are not very active in terms of tweets. However, we see a significant
improvement especially when $\alpha=0.2$ (blue scatter plots). Indeed,
the detected influencers are highly followed as they have about 14k
accumulated followers in total. Besides, the model detected some highly
mentioned and retweeted influencers, especially starting from the
$25^{\textrm{th}}$ detected influencer. Finally, the influence users
selected when $\alpha=0.2$ are more active in terms of tweets than
those selected when $\alpha=0$. 

This first experiment shows the importance of the reliability parameter,
$\alpha$, in detecting influencers with good quality. In fact, we
see that when we consider that all indicators are totally reliable
in characterizing the influence (the case when $\alpha=0$), we notice
that the proposed model detects influencers with very bad quality.
However, when we reduce this reliability ($\alpha=0.2$) we notice
some quality improvement.

\begin{sidewaysfigure}
\begin{centering}
\includegraphics[scale=0.5]{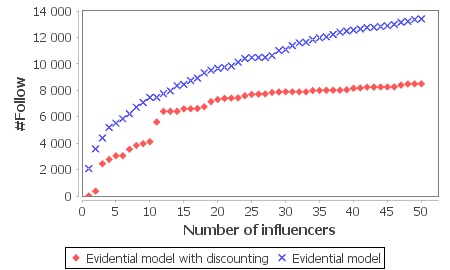} \includegraphics[scale=0.5]{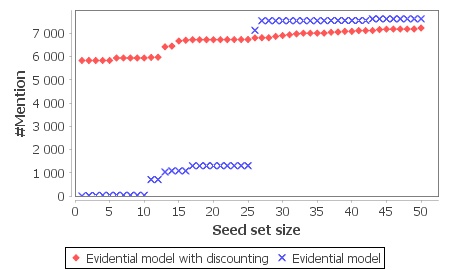}
\par\end{centering}

\begin{centering}
\includegraphics[scale=0.5]{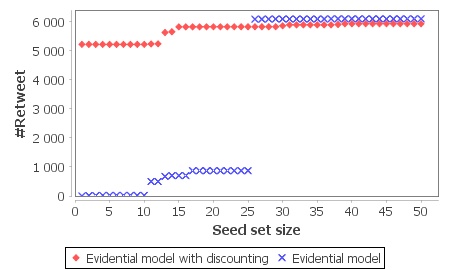} \includegraphics[scale=0.5]{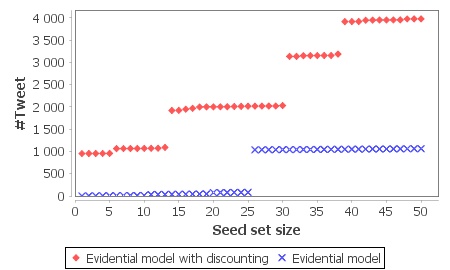}
\par\end{centering}

\caption{Comparison between the proposed reliability-based solution and the
evidential model of \cite{Jendoubi2017}\label{fig:Comparison}}

\end{sidewaysfigure}

In a second experiment, we used the process described in section \ref{sub:Estimating-reliability}
to estimate the reliability of each BBA in the network. Then, each
BBA in our network is discounted using its own estimated reliability
parameter. We note that the parameter $\lambda$ in equation (\ref{eq:discount})
was fixed to $\lambda=5$. To fix this value, we made a set of experiments
with different values of $\lambda$ and the best results are given
with $\lambda=5$. Furthermore, we compare our reliability-based evidential
model (also called evidential model with discounting) to the evidential
model proposed by \cite{Jendoubi2017}. In fact, this last model is
the nearest in its principle to the proposed solution in this paper.
Besides, we fixe the size of the set $S$ of selected influence users
to $50$ influencers. Figure \ref{fig:Comparison} presents a comparison
between the two experimented models in terms of \#Follow, \#Mention,
\#Retweet and \#Tweet (shown in the y-axis of the sub-figures). 

According to Figure \ref{fig:Comparison}, we note that the two experimented
models detect good influencers (shown in the x-axis). However, we
see that the best compromise between the four criteria is given by
the proposed reliability-based evidential model. In terms of accumulated
\#Follow, we notice that the most followed influencers are detected
by the evidential model, also, our reliability-based model detected
highly followed influencers. In terms of \#Mention, we see that the
evidential model starts detecting some mentioned influencers after
detecting about 10 users that are not mentioned. However, the proposed
reliability-based model starts detecting highly mentioned users from
the first detected influencer. Furthermore, we see a similar behavior
in the sub-figure showing the accumulated \#Retweet. Indeed, the proposed
solution detects highly retweeted influencers from the first user.
In the last sub-figure that presents the comparison according to the
accumulated \#Tweet, the best results are those of the proposed reliability-based
model.

This second experiment shows the effectiveness of the proposed reliability-based
influence measure against the evidential influence measure of \cite{Jendoubi2017}.
In fact, the best influence maximization model is always the model
that detects the best influencers at first. Indeed, in an influence
maximization problem we need generally to minimize the number of selected
influencers in order to minimize the cost. For example, this is important
in a viral marketing campaign as it helps the marketer to minimize
the cost of his campaign and to maximize his benefits. Furthermore,
our influence maximization solution gives the best compromise between
the four criteria, \textit{i.e.} \#Follow, \#Mention, \#Retweet and
\#Tweet.

From these experiments we can conclude that the reliability parameter
is important if we want to measure the influence in a social network
through the consideration of several influence indicators. Indeed,
we may have some influence indicators that are more reliable in characterizing
the influence of the others. Furthermore, the proposed solution is
efficient in detecting influencers with a good compromise between
all chosen influence indicators.

\section{Conclusion}

To conclude, this paper introduces a new reliability-based influence
measure. The proposed measure fuses many influence indicators in the
social network. Furthermore, it can be adapted for several social
networks. Another important contribution of the paper is that we consider
the reliability of each chosen influence indicator to characterize
the influence. Indeed, we propose to apply a distance-based operator
that estimates the reliability of each indicator against to the others
and considers the assumption that ``\textit{the farthest from the
others the indicator is, the less reliable it is}''. Besides, we
use the proposed reliability-based measure with an existing influence
maximization model. Finally, we present two experiments that show
the importance of the reliability parameter and the effectiveness
of the reliability-based influence maximization model against the
evidential model of Jendoubi \textit{et al.} \cite{Jendoubi2017}.
Indeed, we obtained a good compromise in the quality of detected influencers
and we had good results according to the four influence criteria,
\textit{i.e.} \#Follow, \#Mention, \#Retweet and \#Tweet.

For future works, we will search to test our influence maximization
solution on other social networks. Then, we will collect more data
from Facebook and Google Plus and we will prove the performance of
the proposed reliability-based influence maximization model. A second
important perspective is about the influence maximization within communities.
In fact, social networks are generally characterized by a community
structure. Then, the main idea is to take profit from this characteristic
and to search to select a minimum number of influence users and to
minimize the time spent to detect them. 

\bibliographystyle{splncs03}
\bibliography{biblio}

\end{document}